\documentclass[10pt,draftcls,twocolumn]{article}
\usepackage{psfrag}
\usepackage{tikz}
\usepackage{pgfplots}
\usepackage{subcaption}
\usetikzlibrary{positioning}
\pgfplotsset{compat=newest} 
\pgfplotsset{plot coordinates/math parser=false} 
\usepackage{amsmath}
\usepackage[dvips]{epsfig}
\usepackage{epsfig}
\usepackage{epstopdf}
\usepackage{setspace}
\usepackage{amsmath,amssymb,bm}
\usepackage{graphics}
\usepackage{booktabs}
\usepackage[pdftex,colorlinks,bookmarks,pdfpagemode=UseOutlines,linkcolor=black,
pagecolor=black,urlcolor=black,citecolor=black,letterpaper,pageanchor=false]{hyperref}
\usepackage[margin=1in]{geometry}
\usepackage[T1]{fontenc}
\usepackage{titlesec}
\usepackage[pagewise,mathlines]{lineno}
\usepackage{longtable}
\usepackage{xfrac}
\usepackage{subcaption}
\usepackage{multirow}
\usepackage{hhline}
\usepackage{titlesec}

\newtheorem{theo}{Theorem}[section]
\newtheorem{lemma}{Lemma}[section]
\newtheorem{df}{Definition}[section]
\newtheorem{cor}{Corollary}[section]
\newtheorem{assump}{Assumption}[section]
\newtheorem{assert}{Assertion}[section]
\newtheorem{remark}{Remark}[section]
\newcommand{\bl}{\begin{lemma}}
\newcommand{\el}{\end{lemma}}
\newcommand{\be}{\begin{equation}}
\newcommand{\ee}{\end{equation}}
\newcommand{\beqn}{\begin{eqnarray}}
\newcommand{\eeqn}{\end{eqnarray}}
\newcommand{\bt}{\begin{theo}}
\newcommand{\et}{\end{theo}}
\newcommand{\bd}{\begin{df}}
\newcommand{\ed}{\end{df}}
\newcommand{\ba}{\begin{assump}}
\newcommand{\ea}{\end{assump}}
\newcommand{\bass}{\begin{assert}}
\newcommand{\eass}{\end{assert}}
\newcommand{\brem}{\begin{remark}}
\newcommand{\erem}{\end{remark}}
\newcommand{\bc}{\begin{cor}}
\newcommand{\ec}{\end{cor}}

\newcommand{\G}{\Gamma}

\long\def\comment#1{}

\titlespacing*{\section}{0pt}{\baselineskip}{0pt}
\title{\LARGE \bf Effect of Adaptive and
Cooperative Adaptive Cruise Control on Throughput of Signalized Arterials}
\author{Armin Askari, Daniel Albarnaz Farias, Alex A. Kurzhanskiy, Pravin Varaiya
\thanks{* This research is funded by California Department of Transportation UCTC Award 65A0529 TO041.}
}

\begin{document}
\maketitle

\begin{abstract}
The paper evaluates 
the influence of the maximum vehicle acceleration  and variable proportions of
ACC/CACC vehicles on the  throughput of an  intersection.
Two cases are studied:
(1) free road downstream of the intersection; and
(2) red light at some distance downstream of the intersection.
Simulation of  a 4-mile stretch of an arterial with 13 signalized intersections is used to evaluate the impact of (C)ACC vehicles on the mean and standard deviation of
travel time as the proportion of (C)ACC vehicles is increased.   The results suggest a very high urban mobility benefit
of (C)ACC vehicles at little or no cost in infrastructure.
\end{abstract}


\section{Introduction}\label{sec_introrev1}
The capacity of a freeway will increase if vehicles traveling at the speed limit could do so at much shorter
headway, since the flow is just the product of speed and density, which is just the inverse of the
headway.  Adaptive cruise control (ACC) and cooperative adaptive cruise control (CACC) are longitudinal control
technologies that permit headway reduction by a factor of 2 or 3 (compared with manual driving), and thereby increasing freeway capacity by the same factor.  A group of vehicles traveling with short headway is called a platoon.  The earliest  demonstration of
an 8-car platoon of vehicles traveling 16 feet apart at 60 mph, 
was conducted on the I-15 freeway in San Diego in August 1997 (\cite{NAHSRC, shladover-keynote}).  Several other demonstrations have  been
carried out since then, e.g. (\cite{ploeg, cacc-milanes}).  These demonstrations show that headway reduction by a factor of 2 to 3 can be achieved today.

However, increasing the capacity of urban roads by platooning would not increase the throughput of an urban road network whose bottlenecks occur at intersections. For example,   In a standard intersection with 4 approaches and two lanes--one through  and one left-turn lane each with
a capacity of 2000 vph(vehicles per hour) --the total capacity of the roads leading to the intersection is $8 \times 2000 = 16,000$vph .  But the intersection can accommodate only two movements safely at the same time, hence the intersection's capacity is only 4,000 vph or 25 percent of the road capacity.  So increasing the latter
capacity by platooning will do nothing to increase the capacity of the road network.

This observation is at the core of the remarkable study \cite{platoons} to increase the throughput or saturation flow rate of an intersection by having
cars cross the intersection in a platoon of ACC or CACC vehicles. The study concludes: if the saturation flow rates at all intersections in an urban network is increased by a factor $\G$, ``the network can support an increase in demand by the same factor $\G$, with no increase in queuing delay or travel time, and using the same signal control. However, the queues will also grow by the same factor $\G$, so if this leads to a saturation of the links, the improvement in throughput will be sub-linear in $\G$. On the other hand, if the cycle time is reduced, the queues will also be reduced, and this may restore the linear growth in demand.''

The authors stress two limitations of their study, which this paper overcomes.  First, the study assumes that there is a 100 percent penetration of (C)ACC vehicles, whereas the present paper studies the effect of an arbitrary proportion of manual, CC and ACC vehicles.  (The effect on freeway capacity of such
arbitrary proportions is evaluated in \cite{mahmaCVT}.)  ``The second limitation is that in short urban links vehicles will slow down quickly as queues build up. As a result the saturation flow rate at the upstream intersection will be reduced, thereby depriving the system of the full productivity benefit. It is important to investigate this reduction.'' That investigation is carried out in this paper.

Our investigation is carried out on the basis of simulaton results using the Improved Intelligent Driver Model \cite{treiberkesting_book} or IIDM in SUMO.  IIDM
improves upon the IDM model \cite{treiberetal2000} by eliminating some of the latter's unrealistic behaviors. IIDM is reviewed in Section~\ref{sec_cfmrev1}. In 
Section~\ref{sec_intersection_throughput} we discuss intersection throughput for manually driven vehicles, and in Section~\ref{sec_acc_caccrev1}, 
we investigate the increase in  throughput due to (C)ACC.  In Section~\ref{sec_platoon_experimentrev1} we introduce the platoon model
and evaluate ACC and CACC in terms of travel time and network throughput using SUMO simulation~\cite{sumo} of the 4-mile stretch of Colorado Boulevard
/ Huntington Drive arterial with 13 signalized intersections in Arcadia, Southern California. Section~\ref{sec_conclusionrev1} concludes the paper.

\section{Car Following Model}\label{sec_cfmrev1}
We start with the notation  in Table~\ref{tab-micro-notation},
and the default parameter values that we will use.
The state  equations for the IIDM car-following model are:
\begin{eqnarray}
v(t+\Delta t) & = & v(t) + a(t)\Delta t ; \label{eq_vehicle_speed}\\
x(t+\Delta t) & = & x(t) + v(t)\Delta t + \frac{a(t)\Delta t^2}{2}.
\label{eq_vehicle_position}
\end{eqnarray}
The critical control variable is acceleration $a(t)$, set by
\begin{eqnarray}
&& a(t) = \nonumber \\
&& \left\{\begin{array}{l}
a_{\max}\left(1-\left(\frac{g_d(t)}{g(t)}\right)^{\delta_1}\right),
\mbox{ if } \frac{g_d(t)}{g(t)} > 1;\\
a^\ast(t)\left(1-
\left(\frac{g_d(t)}{g(t)}\right)^{\frac{\delta_1 a_{\max}}{a^\ast(t)}}\right),
\mbox{ otherwise,}
\end{array}\right.
\label{eq_iidm}
\end{eqnarray}
where
\begin{eqnarray}
a^\ast(t) & = & a_{\max}\left(1-
\left(\frac{v(t)}{v_{\max}}\right)^{\delta_2}\right),
\label{eq_iidm_afree}\\
g_d(t) & = & g_{\min} + \max\Big{\{}0, \nonumber\\
&&  v(t)\tau +
\frac{v(t)(v(t)-v_l(t))}{2\sqrt{a_{\max}b}}\Big{\}},\label{eq_iidm_desired_gap}
\end{eqnarray}
and $\delta_1, \delta_2$ are  fixed positive parameters.
We use $\delta_1=4$, $\delta_2=8$.
The equilibrium speed ($a(t)=0$)
is given by $v(t) = v_l(t) = v_{\max}$
and $g(t) = g_{\min} + v(t)\tau$, and then
the equilibrium headway is
\be
\theta_e = \tau + \frac{g_{\min} + l}{v_{\max}}.
\label{eq_min_headway}
\ee
The default values from Table~\ref{tab-micro-notation} give $\theta_e=2.5$ s,
which translates to $f_e=1/\theta_e=0.4$ vehicles per second, 24 vehicles per minute,
or 1440 vph. Empirical estimates of throughput vary between 1200 and 1900 vph.

\vspace*{-5pt}
\begin{table}[h]
\begin{center}
\begin{tabular}{|l|l|l|}
\hline
Symbol & Description &  \begin{tabular}{@{}l@{}}Default\\ 
value\end{tabular}\\
\hline
$l$ & Vehicle length. & $5$ m\\
$t$ & Time. & \\
$\Delta t$ & Model time step. & $0.05$ s\\
$x(t)$ & Vehicle position. &\\
$x_l(t)$ & Position  of the lead vehicle.&\\
$v_{\max}$ & Speed limit. & $20$ m/s\\
$v(t)$ & Vehicle speed. &\\
$v_l(t)$ & Speed of the leader. &\\
$a(t)$ & Vehicle acceleration. &\\
$a_{\max}$ & Maximal vehicle acceleration. & $1.5 \mbox{ m/s}^2$ \\
$b$ & Desired vehicle deceleration. & $2 \mbox{ m/s}^2$\\
$g(t)$ & \begin{tabular}{@{}l@{}}Gap: distance from the front\\
of the vehicle to the tail of the\\
leader,
$g(t)=x_l(t)-x(t)-l$.\end{tabular} &\\
$g_{\min}$ & Minimal admissible gap. & $4$ m\\
$g_d(t)$ & Desired gap. & \\
$\tau$ & Reaction time. & $2.05$ s\\
$\theta(t)$ & Headway: $\theta(t) = \frac{x_l(t)-x(t)}{v(t)}$. &\\
$f(t)$ & Vehicle flow: $f(t) = \frac{1}{\theta(t)}$. &\\
\hline
\end{tabular}
\end{center}
\caption{Notation summary.}
\label{tab-micro-notation}
\end{table}

\section{Intersection Throughput}\label{sec_intersection_throughput}
\begin{figure}
\centering
\includegraphics[width=3.2in]{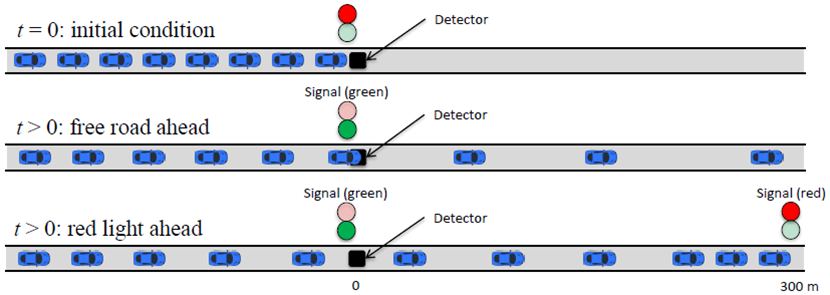}
\caption{Signal turns green at time $t=0$, and vehicles start
moving.
In the first experiment, the first vehicle has free road ahead.
In the second experiment, the first vehicle encounters red light
at the next intersection, 300 meters downstream.}
\label{fig-basic-experiment}
\end{figure}

We analyze the urban arterial of Fig.~\ref{fig-basic-experiment} with infinitely many vehicles in the queue with  minimal gap between them.
At time $t=0$, the light turns green, and vehicles are released. We  discuss two experiments --- one with free road ahead, and the second  with a red light 300 m 
downstream at the second intersection.  This distance accommodates 33 vehicles in the queue
(see Table~\ref{tab-micro-notation} for default values of the car length $l$   and the minimal gap $g_{\min}$), which exceeds the number that
can be sent from the first signal in one minute.

\begin{figure}
\centering
\includegraphics[width=3.2in]{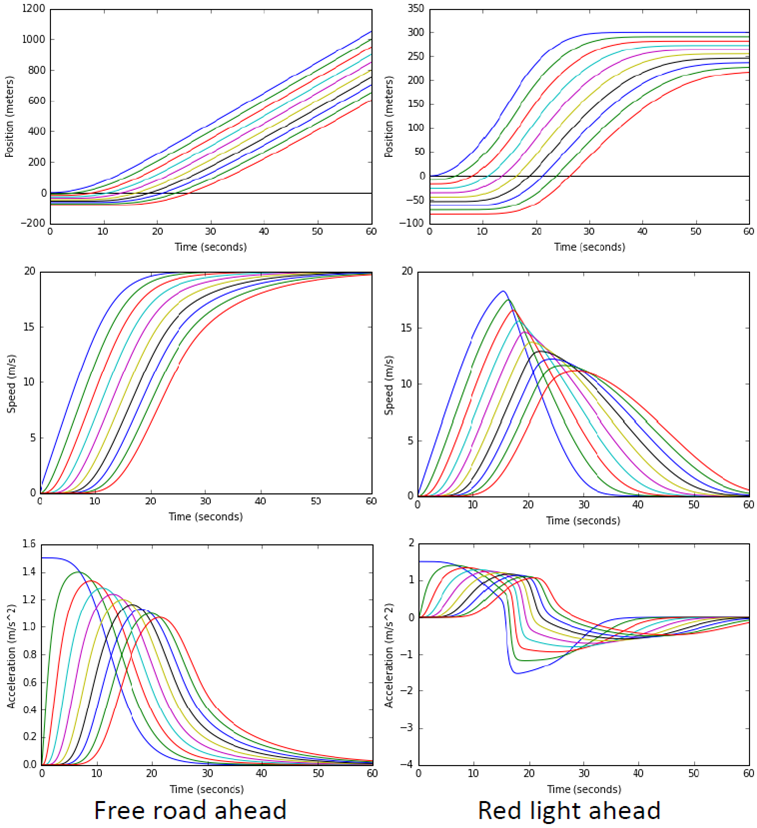}
\caption{Vehicle trajectories, speeds and accelerations:
experiment with free road ahead (left); and
experiment with red light ahead (right).}
\label{fig-trajectories}
\end{figure}

Fig.~\ref{fig-trajectories} displays trajectories, speeds and accelerations of the first ten vehicles from the queue  for these two experiments.
The signal is located at position 0 and indicated by the horizontal black line in the top three trajectory plots.
The leader of the first vehicle is infinitely far ahead.  In the experiment with free road downstream, it accelerates with $a^\ast(t)$
from equation~\eqref{eq_iidm_afree}, asymptotically approaching the maximum speed.
In the second experiment with the red light  at position $x_s=300$, a``blocking vehicle'' is placed at position
$ x_b = x_s + g_{\min} + l $ = 300 + 4 + 5 = 309,
with velocity $v_b=0$, which causes the first vehicle to stop at  $x_s$ to maintain the minimal gap.

\begin{figure}[h]
\centering
\includegraphics[width=3.2in]{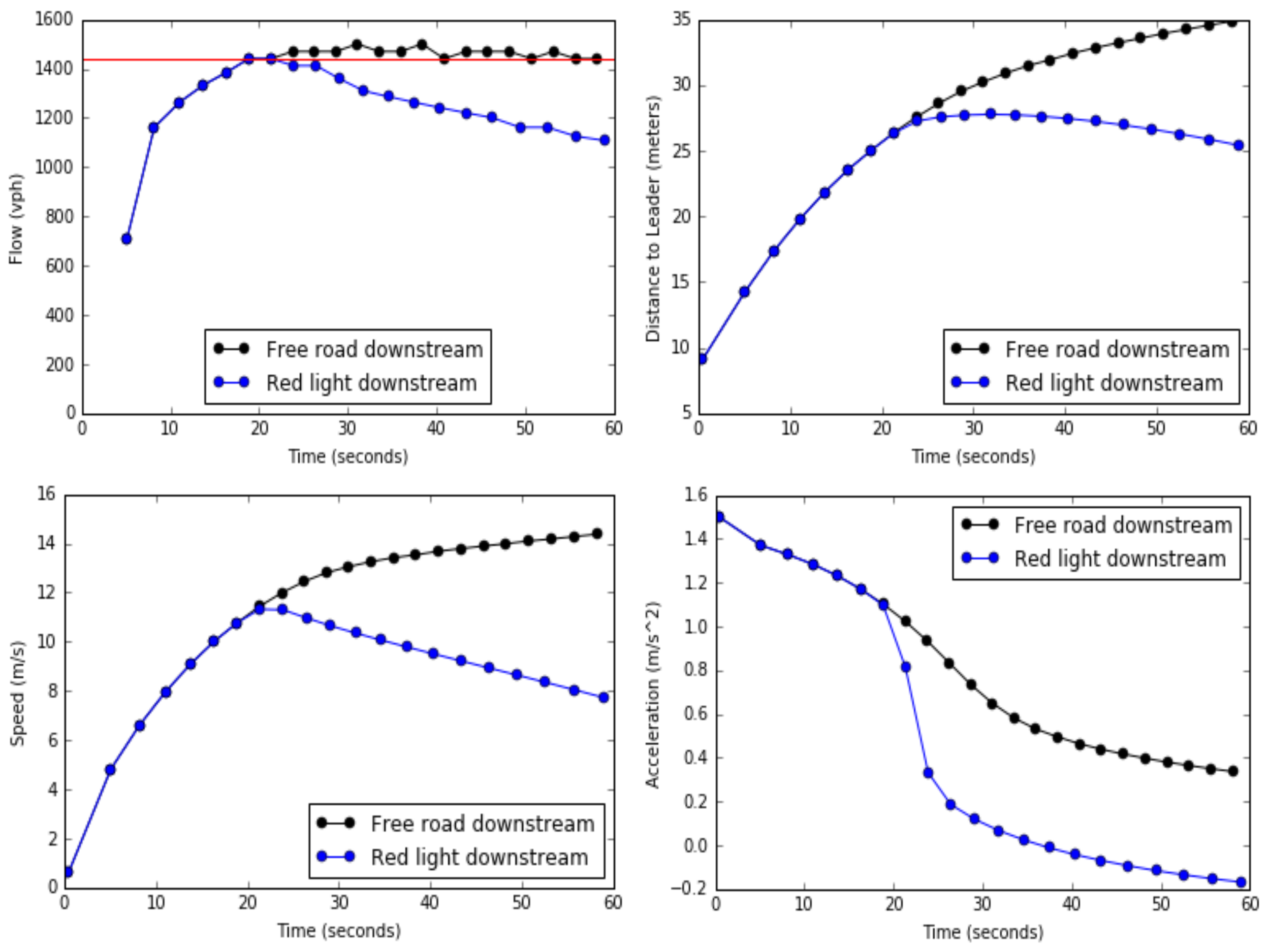}
\caption{Experiment with a free road ahead: comparison of 
measurements of flow, distance to leader, speed and acceleration
at the detector location between the three car following models.}
\label{fig-model-comparison}
\end{figure}

As a queue builds up in front of  the red light, vehicles slow down, and thereby 
reduce the vehicle flow though the
first intersection.
In the experiment with the red light downstream we study the impact of this
braking effect on the throughput of the first intersection.  (This is the second limitation noted in \cite{platoons}.)

The most interesting for the intersection throughput assessment is the traffic behavior at the stop bar location indicated in Fig.~\ref{fig-basic-experiment}.
Fig.~\ref{fig-model-comparison} presents the  measurements obtained from this detector, with each dot corresponding to a vehicle passing the detector.
The (instantaneous) flow (top left) is computed for a vehicle passing the detector based on the
time elapsed after the previously detected vehicle, taken as a headway $\theta(t)$, which is then inverted ($f(t) = 1/\theta(t)$)
and expressed in vph. The red horizontal line corresponds to the equilibrium flow of 1440 vph in our case, when vehicles move at maximal speed.
The gap (top right) between vehicles as well as the speed (bottom left) are monotonically increasing, while acceleration (bottom right) is monotonically decreasing.

In the case of free road downstream, 23 vehicles cross the intersection during the first minute, slightly below the equilibrium flow
of 24 vehicles per minute. The IIDM control can be tuned to a more aggressive behavior by increasing
its parameters $\delta_1,\delta_2$.  In the case of red light downstream, the flow is reduced to 21 vehicles
during the first minute as the result of braking propagation.

\begin{figure}[h]
\centering
\includegraphics[width=3.2in]{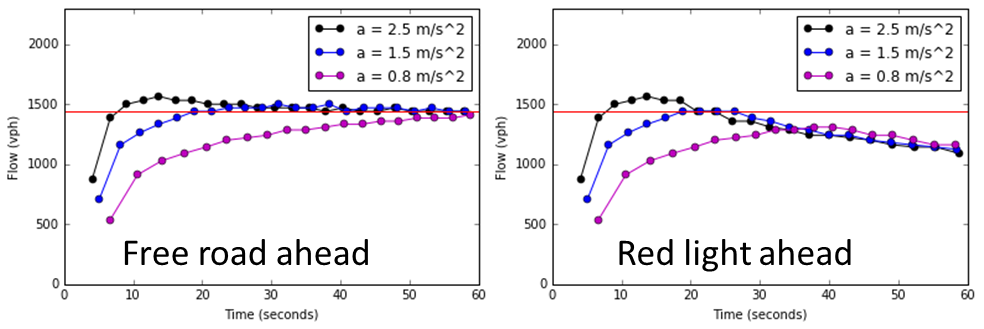}
\caption{Comparing flows for different values of $a_{\max}$
for  experiments  with the free road (left), and
with the red light ahead (right).}
\label{fig-acceleration-comparison}
\end{figure}

We now examine how throughput at the first intersection in our
two previous experiments depends on the maximal acceleration $a_{\max}$.
We repeat both experiments, with the free road and with the red light ahead,
for three different values of $a_{\max}$: $0.8$, $1.5$ (our default) and  $2.5 \mbox{ m/s}^2$.
Fig.~\ref{fig-acceleration-comparison} presents the  measurements obtained at the stop bar location for the cases of free road (left)
and red light downstream (right).
Table~\ref{tab-model-throughput} summarizes the throughput results
for different acceleration values.  The results for this `base' case will be compared with the increased throughput using (C)ACC.

\vspace*{-5pt}
\begin{table}[h]
\begin{center}
\begin{tabular}{|c|c|c|}
\hline
Value of $a_{\max}$ & Type of experiment & IIDM \\
\hline
\multirow{2}{*}{$0.8\mbox{ m/s}^2$} &
free road ahead & 20 \\
 & red light ahead & 19 \\
\hline
\multirow{2}{*}{$1.5\mbox{ m/s}^2$} &
free road ahead & 23 \\
& red light ahead & 21 \\
\hline
\multirow{2}{*}{$2.5\mbox{ m/s}^2$} &
free road ahead & {\bf 24} \\
& red light ahead & 22 \\
\hline
\end{tabular}
\end{center}
\caption{Summary of  experiments: intersection
throughput in vehicles per minute.
Values equal to the equilibrium flow are  in bold.}
\label{tab-model-throughput}
\end{table}

\section{Effect of ACC and CACC}\label{sec_acc_caccrev1}
We will now explore the impact of ACC and CACC vehicles on intersection
throughput.
To do that, we repeat two experiments described in
Section~\ref{sec_intersection_throughput}  but this time,  with a mix of ordinary 
(manually driven), ACC and CACC vehicles.
Values of car following parameters for the three vehicle types
are given in Table~\ref{tab-params-per-vehicle-type}.
As one can see, ACC and CACC vehicles can maintain shorter distances
to the vehicle in front.
\begin{table}[h]
\begin{center}
\begin{tabular}{|l|c|c|}
\hline
Vehicle type & $\tau$ (seconds) & $g_{\min}$ (meters)\\
\hline
Ordinary & 2.05 & 4 \\
ACC-enabled & 1.1 & 3 \\
CACC-enabled & 0.8 & 3 \\
\hline
\end{tabular}
\end{center}
\caption{Values of reaction time $\tau$ and minimal gap $g_{\min}$
for ordinary, ACC- and CACC-enabled vehicles.}
\label{tab-params-per-vehicle-type}
\end{table}

We assume that the ACC vehicle has the same car following model as that of the
ordinary vehicle, but with different $\tau$ and $g_{\min}$.  A
CACC vehicle behaves like an ACC vehicle if
it follows an ordinary vehicle, but if it has another CACC car in front,
it assumes different car following behavior, which we call
\emph{CACC car-following model}.

Let $a_{IIDM}(t)$ denote the acceleration function defined by~\eqref{eq_iidm}.
Define the \emph{constant-acceleration heuristic (CAH)} acceleration 
function~\cite{treiberkesting_book}:
\begin{eqnarray}
&& a_{CAH}(t) = \nonumber \\
&& \left\{\begin{array}{l}
\frac{v^2(t)\bar{a}_l(t)}{v_l^2(t)-2(x_l(t)-x(t)-l)\bar{a}_l(t)}, \\
\mbox{ if } \frac{v_l(t)(v(t) - v_l(t))}{2 \bar{a}_l(t)}\leq (x(t)+l-x_l(t));\\
\bar{a}_l(t) - \frac{(v(t)-v_l(t))^2\Theta(v(t)-v_l(t))}{2(x_l(t)-x(t)-l)},
\mbox{ otherwise,}
\end{array}\right.
\label{eq_a_cah}
\end{eqnarray}
where
\[ \bar{a}_l(t) = \min\left\{\dot{v}_l(t), a_{\max}\right\}, \]
and
\[
\Theta(z) = \left\{\begin{array}{ll}
1, & \mbox{ if } z \geq 0,\\
0, & \mbox{ otherwise}.
\end{array}\right.
\]
We now specify the CACC car following
model~\cite{treiberkesting_book}:
\begin{eqnarray}
&& a_{CACC}(t) = \nonumber \\
&& \left\{\begin{array}{ll}
a_{IIDM}(t), \mbox{ if } a_{CAH}(t) \leq a_{IIDM}(t), \\
a_{CAH}(t) + \\
b\tanh\left(\frac{a_{IIDM}(t)-a_{CAH}(t)}{b}\right),
\mbox{ otherwise.}
\end{array}\right.
\label{eq_cacc_model}
\end{eqnarray}

As before, we run the free road and the red light downstream experiments.
For both scenarios we compute the intersection throughput,
when portion of ACC (CACC) vehicles in the initial queue.
Thus we evaluate 24 cases, each defined by:
(1) experiment: free road or red light downstream;
(2) ACC or CACC; and
(3) percentage of ACC (CACC): 10,
25, 50, 75, 90 and 100\%.

\begin{figure}[h]
\centering
\includegraphics[width=3.2in]{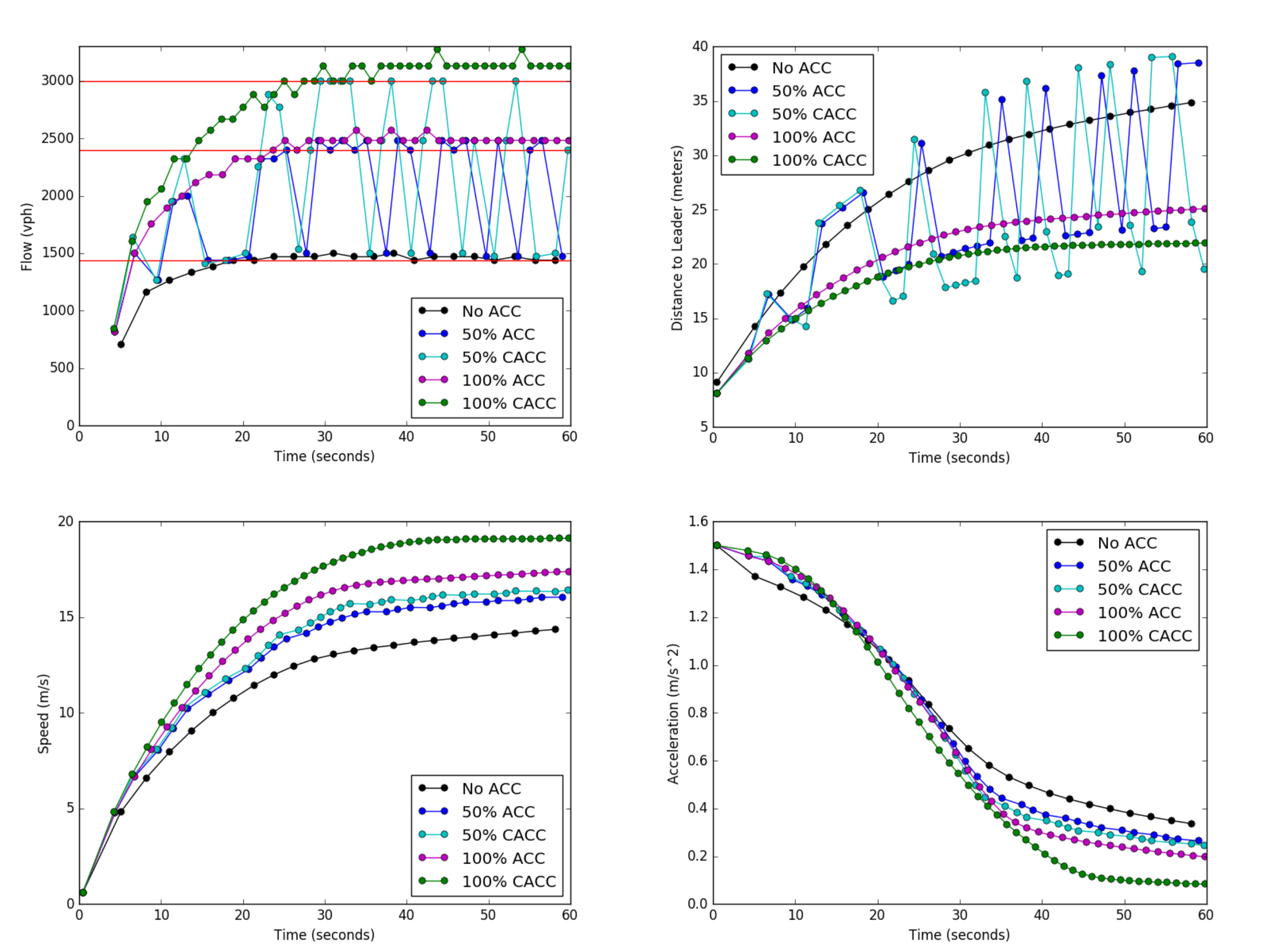}
\caption{Case with free road ahead: comparison of  measurements of flow,
distance to leader, speed, and acceleration at the detector location
for different proportions of ACC/CACC traffic.}
\label{fig-acc-portion-free}
\end{figure}

\begin{figure}[h]
\centering
\includegraphics[width=3.2in]{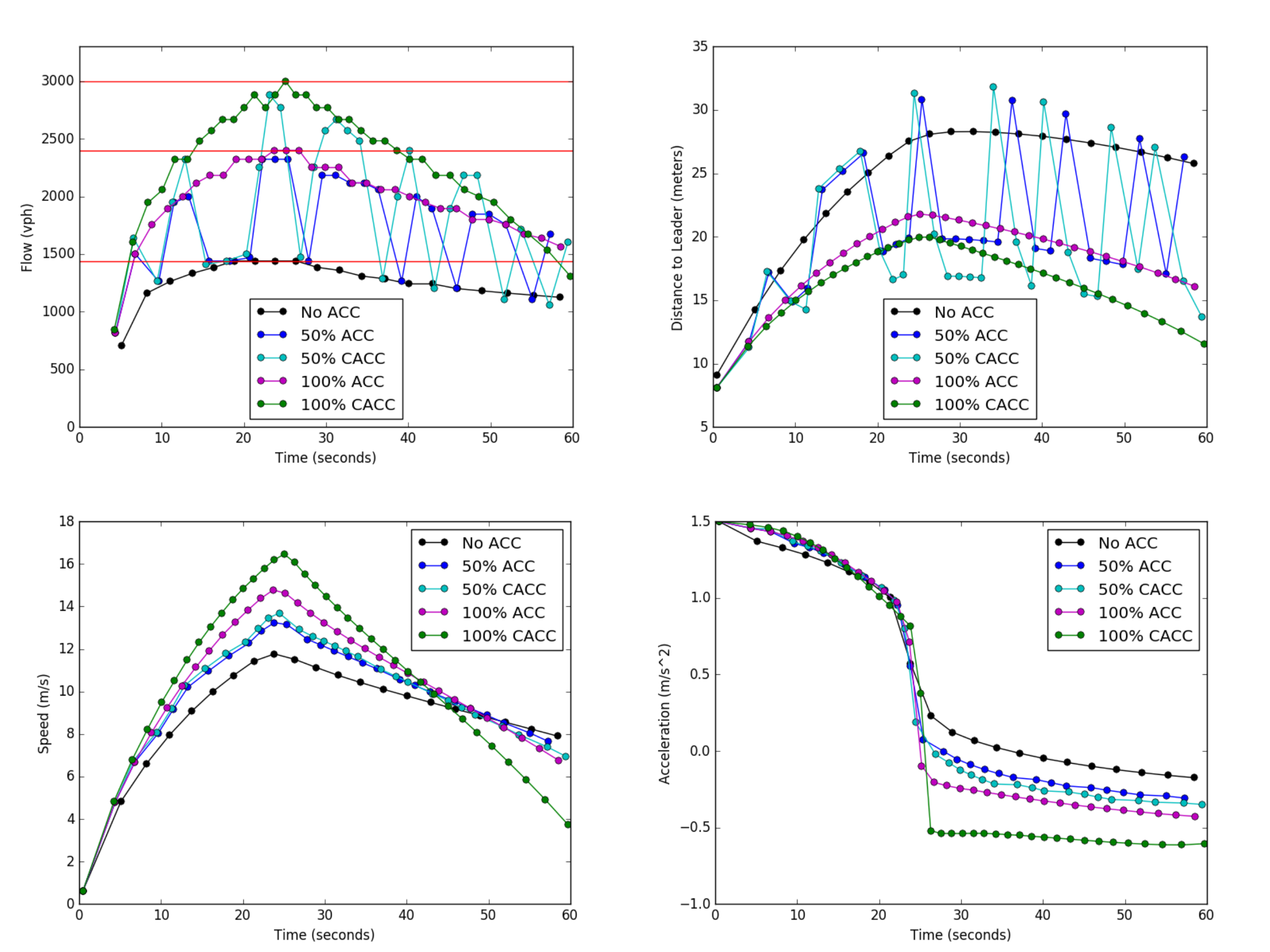}
\caption{Case with red light downstream: comparison of point measurements of flow,
distance to leader, speed and acceleration at the detector location
for different proportions of ACC/CACC traffic.}
\label{fig-acc-portion-stop}
\end{figure}

Fig.~\ref{fig-acc-portion-free} and~\ref{fig-acc-portion-stop} compare
flows, gaps, speeds and acceleration obtained at the detector location
for 0, 50 and 100\% ACC (CACC) penetration rate.
Three horizontal red lines on flow plots in both figures correspond
to equilibrium flows with 0\% ACC (CACC), with 100\% ACC and with
100\% CACC penetration rate.
These flows are computed as $3600/\theta_e$, where $\theta_e$ is given
by~\eqref{eq_min_headway} with $\tau$ and $g_{\min}$ from
Table~\ref{tab-params-per-vehicle-type}, yielding 1440, 2400 and 3000
vehicles per hour respectively.
In the flow and distance to leader plots, notice how 50\% ACC plots
jump between the no ACC and 100\% ACC plots: for an ordinary vehicles
it is similar to the no ACC curve, and for an ACC vehicle it is
similar to 100\% ACC curve.
50\% CACC curves in the same plots jump between three curves --- no ACC,
100\% ACC and 10\% CACC.
This is because a CACC vehicle following an ordinary one behaves
like ACC vehicle. 

For an ACC (CACC) penetration rate less than 100\%, the intersection
throughput is sensitive to the distribution of ACC (CACC) vehicles
in the initial queue.
For example, if 25\% all vehicles in the initial queue are ACC-enabled,
and all of them are concentrated at the head of the queue, we would get
higher vehicle count at the detector location after one minute, than
we would with 50\% ACC penetration rate when all ACC-enabled vehicles are
concentrated at the tail of the queue.
In another example, 50\% CACC penetration rate would not produce any gain
over 50\% ACC penetration rate, if ordinary and ACC/CACC vehicles were
interleaved--- one ordinary, one ACC/CACC, one ordinary, and so on --- since an
CACC achieves a shorter headway than  ACC only when it has other CACC vehicles directly in front.

\begin{figure}[h]
\centering
\includegraphics[width=3.2in]{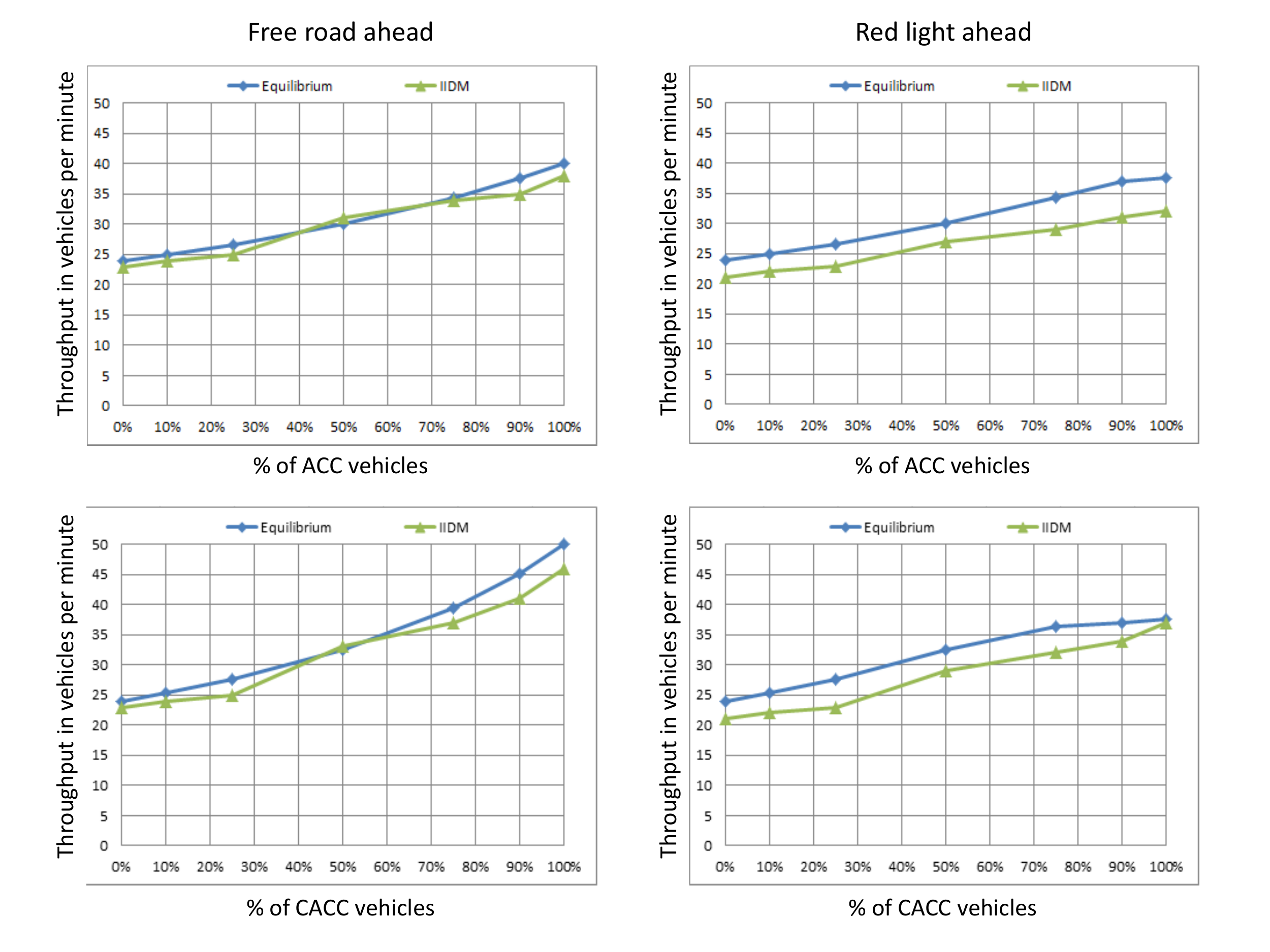}
\caption{Intersection throughput as a function ACC (top) or CACC (bottom)
portion of traffic computed with IIDM for free
road downstream (left) and red light downstream (right).}
\label{fig-acc-cacc-per-model}
\end{figure}

To mitigate this ACC (CACC) distribution bias, in each case with
ACC (CACC) penetration rate less than 100\%, we run 100 random one-minute simulations
of the three car-following models, record vehicle counts at the
detector location, and plot the median vehicle count.
For 100\% penetration rate the ACC (CACC) distribution is trivial, and hence,
a single simulation for each case is enough.
The intersection throughput results for all the 24 cases,
together with throughput values from Table~\ref{tab-model-throughput} obtained
for 0 ACC (CACC) penetration rate, are presented as the four plots in
Fig.~\ref{fig-acc-cacc-per-model}.

Note that in each of the four plots in Fig.~\ref{fig-acc-cacc-per-model},
there is a curve corresponding to the \emph{equilibrium} traffic flow.
These equilibrium curves are computed as follows.
Denote by $\lambda\in[0,1]$ the fraction of ACC (CACC) vehicles in the initial queue, and by
$\tau^{[C]ACC}$ and $g_{\min}^{[C]ACC}$ the reaction time and minimal
gap for ACC (CACC) vehicles, whose values are given in
Table~\ref{tab-params-per-vehicle-type}.
The average headway in the equilibrium state is obtained by modifying
expression~\eqref{eq_min_headway}:
\begin{eqnarray}
\theta(\lambda) & = & \lambda \tau^{[C]ACC} + (1-\lambda)\tau + \nonumber \\
& & \frac{\lambda g_{\min}^{[C]ACC} + (1-\lambda)g_{\min} + l}{v_{\max}}.
\label{eq_modified_headway}
\end{eqnarray}
Then, the equilibrium flow in vehicles per minute is given by:
\be
f(\lambda) = 60/\theta(\lambda).
\label{eq_modified_max_flow}
\ee
This formula is sufficient for the case when there is a free road ahead.
In the case of a red light downstream, however, we are restricted
by the capacity of the link connecting the two intersections.
To account for that, we modify~\eqref{eq_modified_max_flow} accordingly:
\begin{eqnarray}
f(\lambda) & = & \min\Big{\{}\frac{60}{\theta(\lambda)}, \nonumber \\
& & \frac{k\Delta}{\lambda g_{\min}^{[C]ACC} + (1-\lambda)g_{\min}+l}\Big{\}},
\label{eq_modified_max_flow2}
\end{eqnarray}
where $\Delta$ is the length of the link between the two intersections,
and $k$ is the number of lanes in that link.
In our experiment, $\Delta = 300$, and $k=1$.

\section{Platoons on Road Network}\label{sec_platoon_experimentrev1}
Vehicles equipped with CACC can form platoons. 
With 50\% CACC penetration rate, platoons provide between 24 and 44\%
increase in intersection throughput on average, depending on the proximity
of intersections.

\begin{figure}[h]
\centering
\includegraphics[width=3.2in]{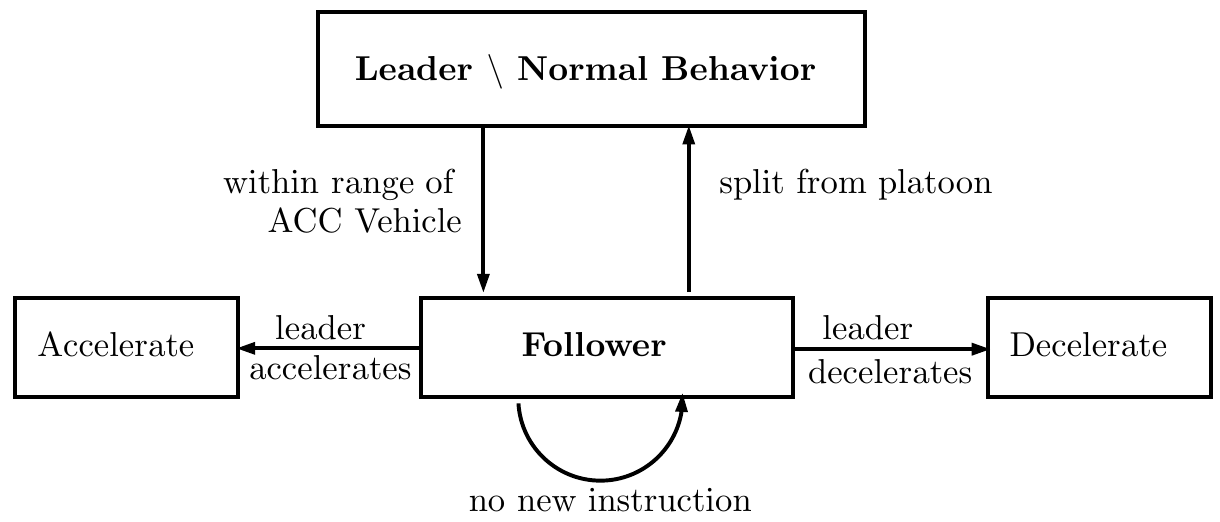}
\caption{State machine of a platoon vehicle.}
\label{fig-state-machine}
\end{figure}

In simulation, platoon management and formation is divided into three phases:
1) Identifying vehicles that can be grouped into platoons;
2) Adjusting parameters of leaders and followers in platoons;
3) performing maintenance on the platoon.
This behavior is modeled by the state machine in Fig.~\ref{fig-state-machine}.

To form a platoon, vehicles must be in sequence
with one another on a given lane.
However, vehicles need not share the same final destination
and are free to switch lanes or leave the platoon if necessary.
If an intermediate vehicle in the platoon changes its route by
making a turn or changing lanes, the platoon splits into two:
one platoon for the vehicles ahead of the intermediate vehicle
and another for all the vehicles behind.

A platoon's lead vehicle has the same properties as ACC vehicles.
An isolated CACC vehicle is a leader of a platoon of size 1.
When a platoon leader comes into range of another CACC vehicle in front,
it joins the platoon becoming a follower.
Followers have reduced headway and travel much closer to one another
than standalone vehicles.
In addition, followers are able to receive information from the leader,
such as to accelerate after a green light at an intersection or to
decelerate approaching an obstacle, e.g. red light, downstream.

Since followers are not bound to the same route as the platoon leader,
they are free to separate.
After leaving the platoon, the headway and acceleration parameters are
restored to their original values.
This can happen for example when the follower changes its route or
becomes separated from the rest of platoon, e.g., due to switching
traffic signal as it crosses the intersection.

\begin{figure}[h]
\centering
\includegraphics[width=3.2in]{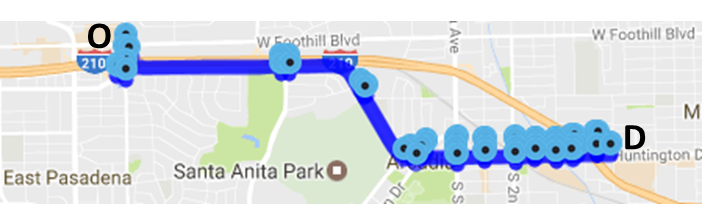}
\caption{Huntington Colorado network modeled in SUMO.}
\label{fig-huntcol}
\end{figure}

To study the impact of platooning, we used a SUMO model of the
4-mile stretch of Colorado Boulevard / Huntington Drive arterial
with 13 signalized intersections in Arcadia, Southern California,
shown in Fig.~\ref{fig-huntcol}.
IIDM and CACC models were implemented in SUMO, and platoon management
and formation were handled via SUMO/TraCI API.
Using real world flow measurements and estimated turn ratios at intersections,
we generated 1 hour  of origin-destination (OD) travel demand data.
Then, we ran a series of simulation varying the fraction of ACC (CACC) vehicles
from 0 to 75\%.
In each simulation two vehicle classes were modeled: ordinary vehicles
and ACC (or CACC) vehicles.
In simulations with CACC vehicles platoons were formed.
The total number of OD pairs in this network is 399.
The same number of vehicles was processed in each simulation.
The rates and locations at which cars were generated were identical
in all scenarios to eliminate the variance in randomly generated routes.
For  cases of 0, 25, 50 and 75 percent  ACC (CACC) penetration rate,
we computed average travel time for the route O$\rightarrow$D, where O and D
identify origin and destination of the  selected west-east route
in Fig.~\ref{fig-huntcol}.
Table~\ref{tab-timesIIDM} lists the mean
travel time (MTT) and its standard deviation (STD),  in seconds.  As expected,
the mean travel time reduces as the fraction of (C)ACC vehicles increases.  Surprisingly the standard deviation also decreases.  Furthermore, the travel time of ordinary vehicles
is also reduced, although that of (C)ACC vehicles is reduced more.

\begin{table}[h]
\centering
\begin{tabular}{|c|c|c|c|c|c|}
\hline
[C]ACC & Vehicle & \multicolumn{2}{|c|}{ACC} & \multicolumn{2}{|c|}{CACC} \\
\cline{3-6}
\% & Class & MTT & STD & MTT & STD \\
\cline{1-6}
0 & ordinary & 653 & 102 & 653 & 102 \\
\cline{1-6}
\multirow{3}{*}{25\%} & ordinary & 640 & 96 & 638 & 96 \\
 & [C]ACC & 605 & 82 & 600 & 76 \\
 & all & 631 & 94 & 629 & 94\\
\cline{1-6}
\multirow{3}{*}{50\%} & ordinary & 583 & 66 & 579 & 60 \\
 & [C]ACC & 583 & 61 & 570 & 64 \\
 & all & 583 & 64 & 575 & 62\\
\cline{1-6}
\multirow{3}{*}{75\%} & ordinary & 595 & 45 & 583 & 41\\
 & [C]ACC & 558 & 58 & 540 & 52\\
 & all & 567 & 57 & 550 & 48\\
\cline{1-6}
\end{tabular}
\caption{Mean travel time (MTT) and standard deviation (STD) in seconds for varying percentage of ACC vehicles on the main arterial of Fig.~\ref{fig-huntcol}.}
\label{tab-timesIIDM}
\end{table}

\section{Conclusion}\label{sec_conclusionrev1}
Presence of (C)ACC-enabled vehicles in the traffic increases the
intersection throughput and reduces travel time at all levels of penetration.
CACC vehicles forming a sequence  increase the throughput of an intersection
significantly more than can be achieved with pure ACC vehicles;  but
 CACC vehicles interleaved with ordinary ones have the same effect
as  ACC vehicles.
Queues in short links do reduce the throughput at an upstream intersection.

In an urban network the presence of (C)ACC vehicles reduces the queues and hence
the time spent at intersections.  As a result ordinary as well as (C)ACC vehicles benefit from
lower average travel times with a lower standard deviation.  The results of this paper together with those  in \cite{platoons}
indicate that (C)ACC can greatly improve urban road mobility at little or no additional infrastructure cost.

\bibliographystyle{IEEEtran}
\bibliography{IEEEabrv,traffic}

\end{document}